# Pattern Formation in a Drying Drop of Starch Gel with Sodium Chloride

Moutushi Dutta Choudhury, Sayanee Jana, Sruti Dutta and Sujata Tarafdar*
Condensed Matter Physics Research Centre
Physics Department, Jadavpur University
Kolkata 700032, India

*Corresponding author, Email: sujata_tarafdar@hotmail.com, Phone: +913324148917, Fax: +913324148917

## Abstract

A drop of aqueous potato starch gel with a certain mole fraction of NaCl, when dried on a glass slide, exhibits strong segregation and intricate pattern formation phenomena. The salt forms radial dendritic crystalline aggregates near the centre, with the starch separating out in the form of a transparent band at the periphery. Photographs and micrographs show a series of concentric rings with different texture.

We report a very simple experiment showing rich and interesting pattern formation. 2 g potato starch (Loba Chemie Pvt. Ltd., Mumbai; M.W $(162.14)_n$) and 1.17 g NaCl are dissolved in 50 ml of deionised distilled water, the mixture is heated to 95°C and stirred till a gel is formed. It is allowed to cool for 1 hr. Then a drop of the gel is deposited on a glass surface and allowed to dry by evaporation (ambient temperature is about 30ºC and humidity varies from 40% to 52%). The dried drop is 1.4cm in diameter. It is observed that the salt accumulates at the centre and crystallizes in dendritic fractal formations. The central circular patch is surrounded by a series of concentric circular bands, each having a different colour and texture. The outermost band is clear and transparent consisting only of starch. The intervening narrow bands with uneven boundary also have a complex structure and composition, yet to be determined precisely.

A close up photograph (taken by Nikon CoolPix L120) of a typical drop is shown in figure 1. Micrographs (Leica DM750) of the different parts, at a magnification of 10x , show an incredibly complex morphology. Representative micrographs of different regions are also shown in figure 1.

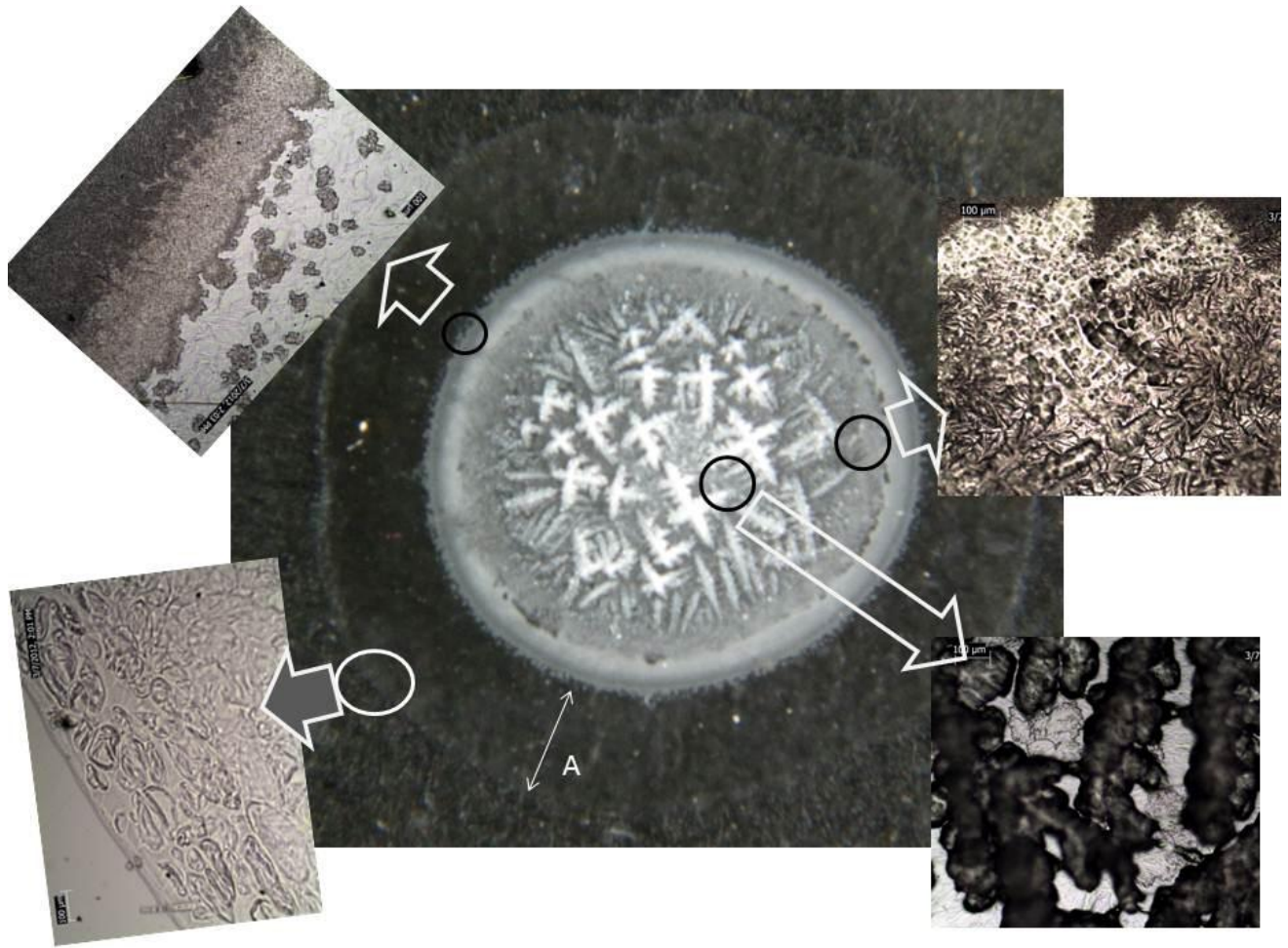

Figure 1. Photograph of a completely dried drop, with micrographs of different regions. The white arrow marked 'A' indicates the outermost transparent band of starch.

Evaporation of fluid drops containing a solute or dispersed granular material has been studied and reported by several workers [1-3]. The most well-known study concerns the 'coffee stain' effect[1]. Here fine coffee granules dispersed in water aggregate along the periphery of the drop forming a ring. Convection currents set up in the fluid due to evaporation from the hemispherical surface are believed to be responsible for this phenomenon. There are reports of a reverse effect where a solvent like octane evaporates leaving the solute to aggregate at the centre of the drop. A strong Marangoni effect is shown to be responsible for this reversal[3]. A polymer like polyethylene oxide containing ammonium perchlorate shows Diffusion-Limited Aggregation (DLA)[4], forming fractal clusters on the surface[5]. Medical science makes use of similar effects in pathological examination of biological fluids[6] and a suspension of salt in albumen[7] shows an effect somewhat similar to the present report. However, in that case all the albumen is reported to move towards the periphery, leaving no albumen near the central region.

Our results resemble closely the results reported by Tarasevich and Pravoslavnova[7], but there is an important difference. In our experiment, the outer band contains *only* starch as revealed in the micrograph, but the central region also has starch in the background of the salt. The starch granules on the periphery can be clearly seen, together with loose strands indicating gel formation. The innermost region shows crystals aggregating in dendritic growth patterns, directed towards the centre. This is in contrast to the polymer films with DLA type growth. It seems that here a strong radial concentration gradient is set up as evaporation takes place along the outer periphery. Probably the concentration gradient, set up in the hemispherical drop drives the crystal aggregation in the sessile drop, rather than diffusion as in the flat film. The crystals allow a slight passage of light under crossed polarizers and thus seem to have some degree of anisotropy, not expected in pure NaCl.

For comparison we have studied drops of only NaCl in water, where a ring of crystals is left along the periphery after drying, similar to the coffee stain phenomenon. The potato starch gel, when dried alone forms a uniform transparent film, somewhat like a dried drop of commercial adhesive. We are at present working on different compositions. The outermost band of starch is found to decrease in width and ultimately vanish, if the ratio of salt: starch is increased. The effect of changing the substrate and solute is also under study. We hope to report these results soon.

Theoretical modelling and analysis[8], as well as experiments of phenomena in drops[9] and films[10] are being pursued by many groups.

We are planning to model the present observations in an attempt to understand the processes which generate this morphology. The points we have noted so far which may help in developing a theoretical model are as follows.

- The initial circular fluid-solid line of contact is pinned to the surface and does not recede during evaporation.
- The shape of the sessile drop is approximately like a spherical cap, with the height decreasing with time.

- The loss in mass (hence volume) of the drop is measured as a function of time. The volume vs. time curve can be fitted with the above assumptions, if a constant rate of evaporation proportional to the surface area of the drop is assumed.

To conclude, the complexity produced by the simple process of mixing salt in starch gel and drying appears quite remarkable and merits further in depth study. This is a preliminary report of this interesting finding.